# High symmetry anthradithiophene molecular packing motifs promote thermally-activated singlet fission

Gina Mayonado[1], Kyle T. Vogt[1], Jonathan D. B. Van Schenck[1], Liangdong Zhu[1], John Anthony[2], Oksana Ostroverkhova[1], Matt W. Graham[1]

1-Department of Physics, Oregon State University, Corvallis, Oregon 97330, USA

2-Department of Chemistry, University of Kentucky, Lexington, Kentucky 40506, USA

**ABSTRACT**

When considering the optimal molecular packing to realize charge multiplication in organic photovoltaic materials, subtle changes in intermolecular charge transfer (CT) coupling can strongly modulate singlet fission. To understand why certain packing morphologies are more conducive to charge multiplication by triplet pair (TT) formation, we measure the diffraction-limited transient absorption (TA) response from four single-crystal functionalized derivatives of fluorinated anthradithiophene: diF R-ADT (R = TES, TSBS, TDMS, TBDMS). diF TES-ADT and diF TDMS-ADT both exhibit 2D brickwork packing structures, diF TSBS-ADT adopts a 1D sandwich-herringbone packing structure, and diF TBDMS-ADT exhibits a 1D twisted-columnar packing structure. When brickwork or twisted-columnar single crystals are resonantly probed parallel to their charge transfer (CT)-axis projections, the TA signal is dominated by a rising component on the picosecond timescale (rate $k_{TT}$), attributed to TT state population. When probed orthogonal to the CT-axis, we instead recover the falling TA kinetics of singlet state depletion at rate $k_A$. The rising to falling rate ratio estimates the TT formation efficiency, $\varepsilon_{TT} = k_{TT} / k_A$ relative to exciton self-trapping. $\varepsilon_{TT}$ ranged from near unity in diF TES-ADT to 84% in diF TDMS-ADT. Interestingly, diF TSBS-ADT crystals only manifest falling kinetics of CT-mediated self-trapping and singlet state depletion. Singlet fission is prohibitive in diF TSBS-ADT crystals owing to its lower symmetry sandwich herringbone packing that leads to $S_1$ to CT-state energy separation that is ~3x larger than in other packings. Collectively, these results highlight optimal packing configurations that either enhance or completely suppress CT-mediated TT-pair formation.

**INTRODUCTION**

Singlet fission, which can generate up to two electron-hole pairs per photon absorbed, has the potential to enhance the efficiency in organic photovoltaics (OPV).[1–4] In well-studied singlet fission materials like TIPS-pentacene, the exoergic conversion of one singlet ($S_1$) excitation to a coupled triplet pair (TT) state is a fast, coherent process that ultimately generates up two free triplets ($T_1$).[5,6] In other systems like tetracene and anthradithiophene (ADT), the $S_1$ to $2T_1$ conversion is much slower owing to an endoergic thermal activation barrier. These endoergic singlet fission processes can conserve energy ($E_{S1} < 2E_{T1}$) and spin only through slower thermal or vibrationally-mediated charge transfer (CT) scattering processes.[7–10] While the underlying mechanisms of endoergic singlet fission have been studied extensively, there are still open questions about how crystal structure promotes intermolecular interactions and maximizes singlet fission yield.[11] In particular, it remains difficult to predict which packing morphologies lead to strongly-coupled CT states that can promote TT formation over competing processes like excimer formation.[12–14] To determine which packing morphologies are most conducive to endoergic singlet fission, we compare the optical spectra and transient absorption (TA) microscopy responses of four single-crystal functionalized derivatives of fluorinated anthradithiophene: diF R-ADT (R = TES, TSBS, TDMS, TBDMS).

Previous studies have established that diF TES-ADT undergoes singlet fission[15–18] but have yet to examine how ADT molecular packing can be modified to promote or suppress the formation of TT states. ADT derivatives all share the same conjugated ring structure that forms the backbone of the molecule with different side groups that attach off the core of the molecule (Figure 1a). These different side groups determine the molecular packing properties of each derivative, which could further affect the rates of singlet fission by altering intermolecular electronic and vibrational coupling. Morphology has been shown to play a significant role in singlet fission in many different materials.[16,19–23] The derivatives investigated in this study include diF TES-ADT, diF TSBS-ADT, diF TDMS-ADT, and diF TBDMS-ADT. As illustrated by the packing morphology cartoons in Figure 1a, diF TES-ADT (*purple*) and diF TDMS-ADT (*green*) both exhibit two-dimensional brickwork packing structures, diF TSBS-ADT (*blue*) has a one-dimensional sandwich-herringbone packing structure, and diF TBDMS-ADT (*red*) has a one-dimensional twisted columnar packing structure. The effects of molecular packing structure on the optical properties of ADT crystals have been explored in detail[24–26] and are summarized in Figure 1a, which shows the steady-state absorption and photoluminescence (PL, *dashed lines*) spectra of the four different derivatives. Figure 1a shows that molecular packing leads to distinct intermolecular coupling as seen by a decreasing spectral Stokes shifts from 0.13 eV in the brickwork packing motifs to 0.058 eV in the less-ordered sandwich herringbone packing motif. When the crystals are dilutely dissolved in dichlorobenzene (< 0.001 mM), the intermolecular coupling vanishes and the solvation-induced spectral Stokes shift is approximately the same for all ADT derivatives and minimal (typically < 5nm). These molecular packing properties make ADT derivatives ideally suited for studying the effect of morphology variation on singlet fission as an intermolecular process.

Studies of singlet fission materials pentacene, tetracene, hexacene, and rubrene have previously used polarization-dependent TA measurements to identify contributions from singlet and triplet excitons.[27–30] As in many such organic crystals, the packing morphology of ADT crystals lead to intermolecular CT states and strongly anisotropic electronic and optical properties. The different transition dipole moments within the ADT crystal structures are preferentially aligned along different axes in real space (see Figure 2b), with the $S_0S_1$ transition dipole oriented at 90° relative to the long crystal axis. Optical absorption in organic crystals is highly anisotropic not only for the $S_0S_1$ transition but also for transitions from excited states.[19] The anisotropic nature of these crystals can be utilized to isolate different ESA transitions in TA measurements, including the formation of stable TT states.

For singlet fission to occur, materials must meet the energetic requirements to convert a singlet exciton into two triplet excitons. When assisted by thermal and vibrational processes, singlet fission in endoergic systems can overcome the energetic barrier and proceed to form TT states.[31,32] Many weakly endoergic systems, including anthracene, tetracene, and diF TES-ADT, still efficiently undergo singlet fission.[17,33–35] diF TES-ADT has previously been categorized as a weakly endoergic system, with literature reporting the 2×$T_1$ energy, $2E_{T1}$, being between 25 and 60 meV higher than the $S_1$ energy, $E_{S1}$.[15,36] The Stokes shift, $E_{Stokes}$ in Figure 1a is used to estimate the lowest singlet Frenkel energy ($E_{S1}$ – ½ $E_{Stokes}$) giving values that range from in 2.19 eV in diF TES-ADT to 2.33 eV in diF TSBS-ADT. Further supporting diF TES-ADT as a moderately endoergic system, DFT calculations give $2E_{T1}$ = 2.20 eV, suggesting only a room temperature-range thermal activatation ~10-30 meV is required.

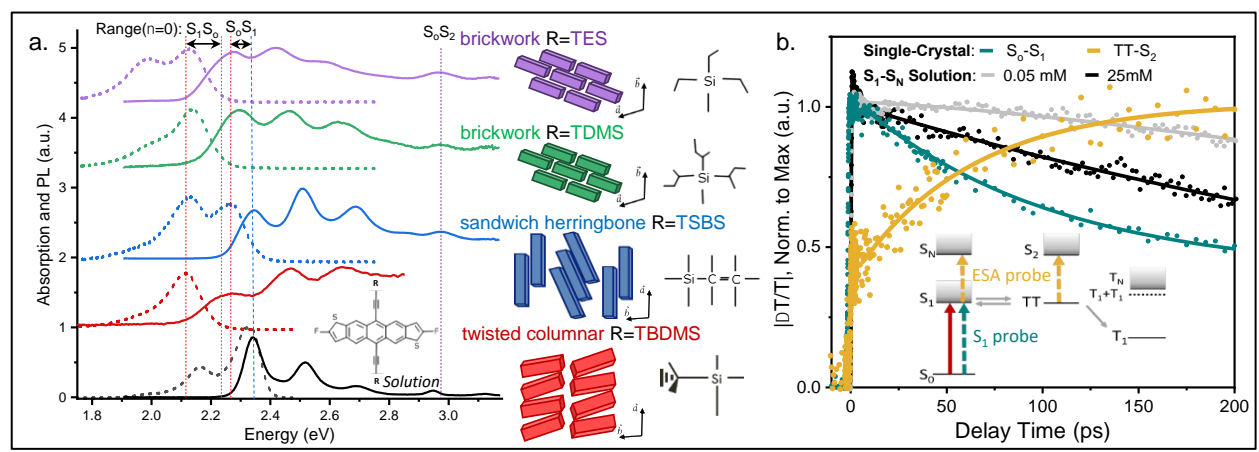

**Figure 1 - Endoergic singlet fission in ADT derivatives. (a)** Absorption (*solid lines*) and photoluminescence (*dashed lines*) spectra of four diF R-ADT derivatives and solution. The cartoon, on right shows the top-down crystal packing morphology obtained by functionalizing the diF R-ADT molecule with the R-groups shown. Polarization is 90° relative to the long crystal axis. **(b)** TA microscopy of a diF TES-ADT crystal gives falling kinetics using a 2.3 eV $S_oS_1$ probe (*aqua line*), but the matching rising kinetics are recovered when probed at 1.26 eV (*yellow line*), suggesting TT-pair formation. While the solution-phase TA relaxes more slowly, it accelerates as concentration increases from 0.05 mM (*grey curve*) to 25 mM (*black curve*), suggestive of aggregation-induced singlet fission. *(Inset)* A basic energy-level diagram illustrates how the probe pulse energy was varied to resolve either falling singlet ground state bleach or rising TT-pair ESA responses.

Compared to the faster, direct singlet fission mechanisms of exoergic systems like TIPS-pentacene, endoergic systems like ADT are characterized by slower, CT-mediated singlet fission dynamics. A representative energy diagram of our system is shown in Figure 1b (inset), which illustrates how our experiment pumps the $S_1$ state and then probes both singlet ($S_1 \rightarrow S_n$) and triplet pair (TT $\rightarrow S_2$) excited state transitions. Using ultrafast TA microscopy on single diF TES-ADT crystals, Figure 1b shows the decaying singlet ground state bleach (*aqua line fit*), roughly matches the triplet pair rising ESA (*orange line fit*). We associate these matching rising and falling curves as evidence of delayed TT formation that has been previously reported in this material.[15,16] Figure 1b further plots the $S_1$ relaxation dynamics of a diF TES-ADT crystal alongside diF TES-ADT solutions in dichlorobenzene at various concentrations. In solution-phase, the TA kinetic decays shown for diF TES-ADT are approximately invariant for all four chemical diF R-ADT functionalizations (see supplementary Figure S1). The TA response of diF R-ADT in the most dilute solution decays most slowly without undergoing singlet fission because this system lacks the necessary intermolecular interactions. However, as the diF R-ADT concentration increases, the decay rate accelerates, suggesting that the singlet state is depleted faster by dimer-mediated TT state formation. This introductory result demonstrates the importance of intermolecular coupling to initiate the singlet fission process, which we will explore further by comparing diF R-ADT crystals of different packing motifs.

**METHODS**

Thin crystals of different crystal packing motifs were grown from diF R-ADT functionalized derivatives through a drop-cast method with controlled evaporation. Each functional substituent group on the ADT molecule is shown in Figure 1a, and is synthesized to yield one of three distinct packing motifs that can best be described as brickwork (TES and TDMS groups), twisted columnar (TBDMS group), and sandwich herringbone (TSBS group).[24] As molecular crystals, these substituents critically control packing and intermolecular coupling as indicated by their differing Stokes shifts relative to solution as highlighted in Figure 1a (dashed lines). Crystals were grown on fused substrates upon which a 45 nm layer of silver was deposited to enhance optical signals collected in a reflective geometry. A droplet of a dichlorobenzene solution of each ADT derivative was deposited on the substrate and grown under a slow solvent evaporation method at 3°C in a covered flask. This procedure yielded well-defined crystals that have been characterized using XRD methods (see Supplementary Figure S2). Coupled with the observation of clear crystal boundaries seen in polarization-dependent microscopy (see Supplementary Figure S3), nearly single-crystal regions could be selectively excited using the diffraction-limited optical probes used throughout this study. Spatially resolved optical characterization of absorption, PL and PLE spectra was done via an all-reflective optics microscope system with a 300 - 800 nm scannable Xe arc lamp source (Horiba Fluorolog). Polarization-dependent hyperspectral imaging was accomplished on an electron-multiplying CCD camera (Princeton Instruments ProEM). For all measurements throughout, the diF R-ADT crystals were kept under high-vacuum in an optical cryostat (Advanced Cryonics LT-003M or Quantum Design OptiCool) to prevent photodegradation.

TA dynamics were investigated in these crystals using diffraction-limited TA microscopy. Ultrafast laser pump pulses were tuned to be resonant with the diF R-ADT $S_o$-$S_1$ transitions using either: (1) ~180 fs pulses at 80 MHz from a second-harmonic generation system (SHG, APE HarmoniXX) pumped by a Coherent Ultra Chameleon, or (2) 35 to 95 fs pulses at 10 kHz from a non-collinear optical parametric amplifier (NOPA, ORPHEUS-N, LightConversion). The high repetition rate measurements enabled measurement collection at the lowest photon fluxes, typically < $5 \times 10^{12}$ photons/cm$^2$, where contributions from effects like triplet-triplet annihilation were undetectable. Tunable optical probe pulses were generated from either: (1) a broadly tunable Ti:Sapphire output or (2) a second optical parametric amplifier (LightConversion, ORPHEUS-F). Furthermore, white light supercontinuum probes were used for collecting optical bleach signals near the $S_1$ transition and for all spectrally-resolved TA microscopy. To identify prominent ESA peaks, the probe energy was scanned from 1 eV to 1.45 eV. This near-IR energy probe range was selected to be approximately half the $S_o$-$S_1$ transition energy. This less congested region of the spectrum enables clear identification of the polarization-sensitive singlet fission kinetics. Specifically, the primary ESA signals accessible with these probe energies were either singlet ($S_1 \rightarrow S_n$) or triplet pair-dominated (TT $\rightarrow S_2$).

The TA microscopy method uses collinear pump-probe beamlines raster-scanned by a piezo-scanning mirror in a 4*f*-confocal geometry to direct the beamlines to the diF R-ADT samples held under a high-vacuum environment. Nearly diffraction-limited excitation is achieved with high-NA (>0.4), NIR-compatible objectives with cover glass correction (Olympus 50XIR and Beck 15X reflective objective). Pulse widths are optimized using cross-correlation measured by photoconduction at the sample position. The piezo-scanning beam and sample stages are used to align to a target crystal whose polarization-dependent spectra were measured by hyperspectral microscopy. The pump and probe polarizations are controlled via two $\lambda$/2 plates. All polarization angles are measured with respect to the long crystal axis, which corresponds to the *b*-axis in diF TES-ADT and diF TDMS-ADT, and the *a*-axis in diF TBDMS-ADT and diF TSBS-ADT. Unless specified, the pump polarization was selected to align to the $S_oS_1$ dipole projection along the single-crystal axis of each diF R-ADT crystal. The transient reflection TA microscopy signal is retrieved by TEM-cooled InGaAs photodetector and lock-in detection (Zurich HF-LI) at selected pump beam modulation rates between 1 to 500 kHz. High modulation rates were achieved with an acousto-optical beam deflection with a $TeO_2$ crystal. Since each transition has a different transition dipole orientation, transitions were isolated in TA microscopy measurements by varying the polarization of the probe beam. Solution-based TA measurements were taken in a separate conventional transmissive pump-probe setup with 1 MHz modulation.

**RESULTS**

Figure 2a plots the PL and PLE spectra of diF TES-ADT single crystals. At lower temperatures, the PL emission is dominated by emission from a lower-lying peak at an energy of $2E_{T1} - E_{vib}$. As shown in previous studies, this $2E_{T1} - E_{vib}$ energy corresponds to emission from the TT state that is allowed through a Herzberg-Teller vibrational coupling mechanism.[15,16,24,36] The polarization-dependent PLE spectra of diF TES-ADT crystals in Figure 2a are obtained using transmissive hyperspectral microscopy imaging for 1.93 eV ± 0.04 eV emission. Interestingly, the $S_oS_2$ peak in the PLE spectra is the most strongly anisotropic, being minimized to almost extinction at 0°. However, the monomer $S_oS_2$ selection rules for both anti- and syn- configurations of diF R-ADT predict maximum absorbance occurs closer to 0°, conflicting with the observed maximal absorption at 90°.[24] To resolve, Figure 2a (*inset*) shows the DFT-calculated CT-axis dipole is at nearly 90° in a top-down view of the brickwork structure of diF TES-ADT. Rather than following the monomer selection rules, the $S_oS_2$ transition polarization aligns to the CT-axis, suggesting absorption to an excited state with CT character. Strong CT-character of the $S_oS_2$ transition is further supported by the near-perfect alignment of the 2.99 eV $S_oS_2$ absorption peak across all four derivatives shown in Figure. 1a. The CT-character of the $S_2$ absorption will be critical to understanding the polarization-dependent TT to $S_2$ ESA responses of diF R-ADT single crystallites shown later in Figure 3.

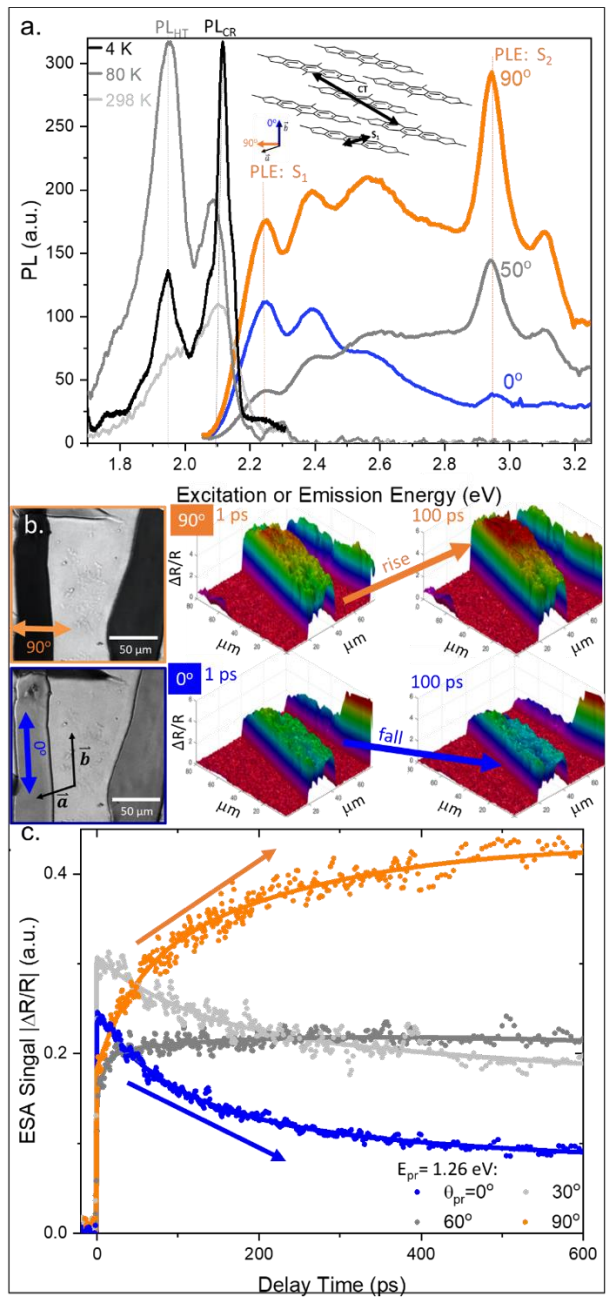

**Figure 2 - Polarization-sensitive TA microscopy. (a)** Temperature-dependent PL spectra of a diF TES-ADT crystal highlight the growth of a low-lying emissive state at 1.95 eV. The PLE spectra of a diF TES-ADT crystal highlight an $S_2$-peak that is essentially switchable with polarization. (*Inset*) Top-down view of ADT molecules stacked in the brickwork packing structure show the calculated dipole orientation of the $S_1$ ground state and the CT state. **(b)** *(left)* Microscopy images of diF TES-ADT crystals for polarization along the short (90°, orange arrow) and long axes (0°, blue arrow) show strong absorption anisotropy. *(right)* TA microscopy snapshots of a movie showing spatial uniform TT formation in a smaller single-crystal region for 90° (*top, rising dynamics*) vs. 0° (*bottom, falling dynamic*) probe polarizations at 1.26 eV. **(c)** Corresponding TA-microscopy relaxation dynamics at 0° (*blue line*), 30° (*light gray*), 60° (*dark gray*), and 90° (*orange*) probe polarizations show that matching rising and falling rates are realized only for the orthogonal polarizations shown.

The hyperspectral microscopy images in Figure 2b show the strong $S_0S_1$ linear absorption anisotropy of nearly single crystals of diF TES-ADT. Figure 2b further plots time-resolved movie snapshots of a ~20 μm wide crystal blow-up of a single crystal region with probe polarizations at 0° (blue arrow) and 90° (orange arrow). These spatially-resolved TA microscopy maps highlight the uniform nature of the rising and falling relaxation kinetics of a diF TES-ADT crystal region. Comparing the 1 ps and 100 ps TA microscopy snapshots of the dynamics, we observe a spatially uniformly rising ESA response for 90° probe polarization and a matching falling ESA response at 0° probe polarization. Such dynamics are difficult to isolate in regions with overlapping domains or larger laser spot sizes because the rising and falling kinetics become convolved due to different dipole orientations. See our Supplementary Video V1 –0 - 800 ps range) for compiled TA microcopy maps that animate the matching rising and falling dynamics summarized in Figure 2b.

Consistently, the most distinct separation of rising and falling TA microscopy kinetics are observed for probe polarizations aligned either parallel or perpendicular to the crystal axes. Figure 2c further includes diF TES-ADT kinetics taken at 30° and 60° intermediate probe polarizations to demonstrate these dynamics are composed of weighted contributions from the kinetic rate processes shown for 0° and 90°. As a result, all other TA pump-probe dynamics are plotted only for probe beams that are polarized either parallel (0°) or perpendicular (90°) to the long axis of the crystals.

For each of the four diF R-ADT packings motifs shown, Figure 3a compares the normalized TA signals taken at 0° (*blue lines*) and 90° (*orange lines*) probe polarizations. In Figures 3ai and ii, the brickwork-packed structures (diF TES-ADT and diF TDMS-ADT) show prominent exponential rises for 90° probe polarizations, oriented perpendicular to the long axis of the crystal. Conversely, for 0° probe polarization, the rising kinetics are replaced by matching falling kinetics. This ~76 ps matching rising and falling time constant is associated with the depletion of the $S_1$ exciton to form a correlated triplet pair state, TT. The linear absorption spectrum for the twisted columnar packing structure, diF TBDMS-ADT, in Figure 3b shows that the maximum $S_0S_2$ resonance is orthogonal to that seen in brickwork packings. Consequently, the TT formation rise was strongest when the probe beam was polarized along the long axis of the crystal (0°, blue line). Unlike the brickwork packings, Figure 3a shows that the ~37 ps rise appears much faster than the 62 ps fall. The fitting parameters in Table 1 account for this by using a biexponential rise to allow for contributions from both coherent vibrational and thermal excitations to overcome the endoergic activation barrier for TT formation. The polarization-dependent absorption spectra shown in Figure 3b support that the distinctive ESA rises shown in Figure 3a occur in diF TES-ADT, diF TDMS-ADT, and diF TBDMS-ADT at whichever polarizations match the dipole of the final $S_2$ absorption.

Most interestingly, the kinetics of the sandwich herringbone packing structure, diF TSBS-ADT, in Figure 3aiii are entirely different. In particular, over a wide range of tunable probe energies (1.15 to 1.4 eV), only falling relaxation kinetics are extracted that are nearly invariant to probe polarization. Moreover, the diF TSBS-ADT relaxation kinetics are consistently faster than solution-phase kinetics (see Figure 1b). Collectively, this is consistent with a related work[24] that suggest this packing favors exciton self-trapping instead over Frenkel emission.

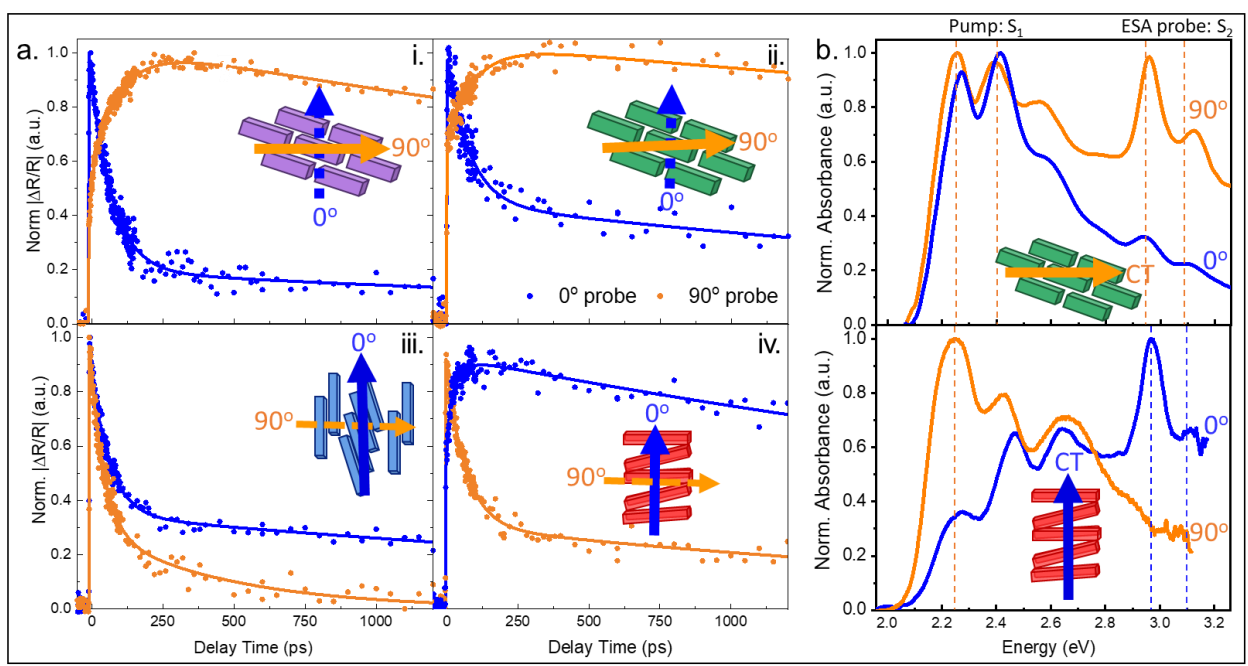

**Figure 3 – diF R-ADT packing morphology-dependent singlet fission. (a)** TA relaxation dynamics of four different molecular crystal packings (shown in insets). Each plot compares ESA dynamics for 0° (blue) vs. 90° (orange) probe beam polarizations. **i.** and **ii.** show TA dynamics of diF TES-ADT and diF TDMS-ADT crystals with a 2.53 eV $S_1$-pump and 1.26 eV $S_2$ region-probe. Both of these brickwork-packed crystals have matching rise/fall times given in Table 1. **iii.** TA dynamics of a diF TSBS-ADT crystal with a sandwich-herringbone packing only shows falling dynamics at all polarizations and probe energies (1.15 to 1.4 eV) investigated. **iv.** The rising/falling TA dynamics shown of diF TBDMS-ADT crystals occur at probe polarization opposite to brickwork packing owing to the twisted columnar packing structure illustrated. **(b)** Normalized absorption spectra of the corresponding diF TDMS-ADT (*top*) and diF TBDMS-ADT (bottom) crystals at 0° (blue) and 90°(orange) polarizations. Rising TA dynamics shown are always best observed for the polarization where the $S_0S_2$ transition peak is maximal.

Table 1 gives the corresponding least-squares deconvolution fitting rates of the plots in Figure 3a for each probe polarization in these diffraction-limited TA microscopy measurements. This preliminary analysis explicitly ignores any sub-picosecond dynamics beyond the instrument response function (IRF), and uses a least-squares biexponential ($\tau_1$, $\tau_2$) fit to capture the two dominant timescales over the 1.2 ns scanning range shown. This measurement employed a low pump fluence of 3.5 × 10$^{12}$ photons/cm$^2$ and was in the strongly linear region of pump-power dependence. Later in Figure 6, further data will show how the faster dynamics (< 200 fs) contribute to morphology-dependent singlet fission and exciton-exciton annihilation in diF R-ADT crystals. Kinetic analysis (see discussion section II) of each time constant permits assignment to the following polarization-dependent ESA processes: (i) $\tau_1$ rise: the TT → S$_2$ ESA is associated with TT formation at the rate k$_{TT}$, (ii) $\tau_1$ fall: the S$_1$ → S$_n$ ESA is primarily associated with the S$_1$ depletion to the TT state and self-trapped excitons (STE) at rate k$_A$, (iii) $\tau_2$ fall: primarily associated with the S$_1$ and TT recombination rates. The rise and fall times of $\tau_1$ match between the Table 1 columns taken for opposite probe polarizations within the stated error. Singlet fission on very similar timescales has been observed in other endoergic systems including tetracene, where the roughly matching rising/falling rates are consistent with TT state formation through a thermally activated endoergic singlet fission barrier.[37]

| diF R-ADT; R= | 0° $\tau_1$ (ps) | 90° $\tau_1$ (ps) | 0° $\tau_2$ (ns) ↓ | 90° $\tau_2$ (ns) ↓ |
|---|---|---|---|---|
| TES brickwork 1 | 76 ±2 ↓ | 74 ±3 ↑ | 3.1 ±1.4 | 7.1 ±1.2 |
| TDMS brickwork 2 | 78 ±9 ↓ | 93 ±9 ↑ | 3.5 ±1.3 | 10 ±2 |
| TBDMS twisted columnar | 9 ↑, 65 ↑ | 62 ±5 ↓ | 3.0 ±0.3 | 2.7 ±0.9 |
| TSBS sandwich herringbone | 64 ±2 ↓ | 49 ±3 ↓ | 3.5 ±0.6 | 0.47 ±0.7 |

**Table 1** Summary of the falling (↓) and rising (↑) $\tau_1$ and $\tau_2$ time constants of probe-polarization dependent TA. The nearly matching rise/fall times in three of the four diF-R-ADT crystal packing are associated with CT-mediated TT-pair state formation, a precursor for singlet fission.

The TA microscopy kinetic traces plotted in Figure 4a show all four possible pump and probe polarization permutations labeled as $\theta_{pump}/\theta_{probe}$. While the probe-beam polarization impact is strong, the ESA kinetic rates are only weakly impacted by changing the pump polarization. The ESA polarization-dependent dynamics are determined by the probe polarization and the coupling strength between its electric field and the available transition dipole moments.[38] The transition dipole moments and orientations in diF R-ADT crystals were investigated using DFT dimer calculations summarized in Figures 4b-c and are further described in Supplementary Table S2 and in J. Van Schenck et al.[24] The DFT-calculated Frenkel (*purple dashed arrow*) and the strongest CT dipole orientations (*red dashed arrow*) are plotted in Figure 4b relative to the crystal plane of light incidence. DFT calculations for diF TBDMS-ADT give two similar CT-axis orientations due to its twisted columnar structure. To compare the calculated dipole orientations against experimental predictions, Figure 4b shows that the solid green arrow aligns to the polarization that maximizes the $S_0S_2$ absorption. This calculated and observed transition dipole moment is always along the axis of the strongest intermolecular coupling in each ADT packing morphology. The approximate agreement of the calculated strongest CT-state dipole orientation and that of the $S_0S_2$ absorption peak suggests that this ~2.99 eV transition is borrowing oscillator strength from the CT-state transition. We further observe the CT-mixed $S_0S_2$ transition dipoles (obtained from the polarization-dependent optical absorption spectra) and the rising TT-formation dynamics (obtained from polarization-dependent ESA) have matching probe polarization across all packing morphologies. The observed probe-beam polarization dependencies align very closely with those predicted by TD-DFT calculated transition dipole moments.

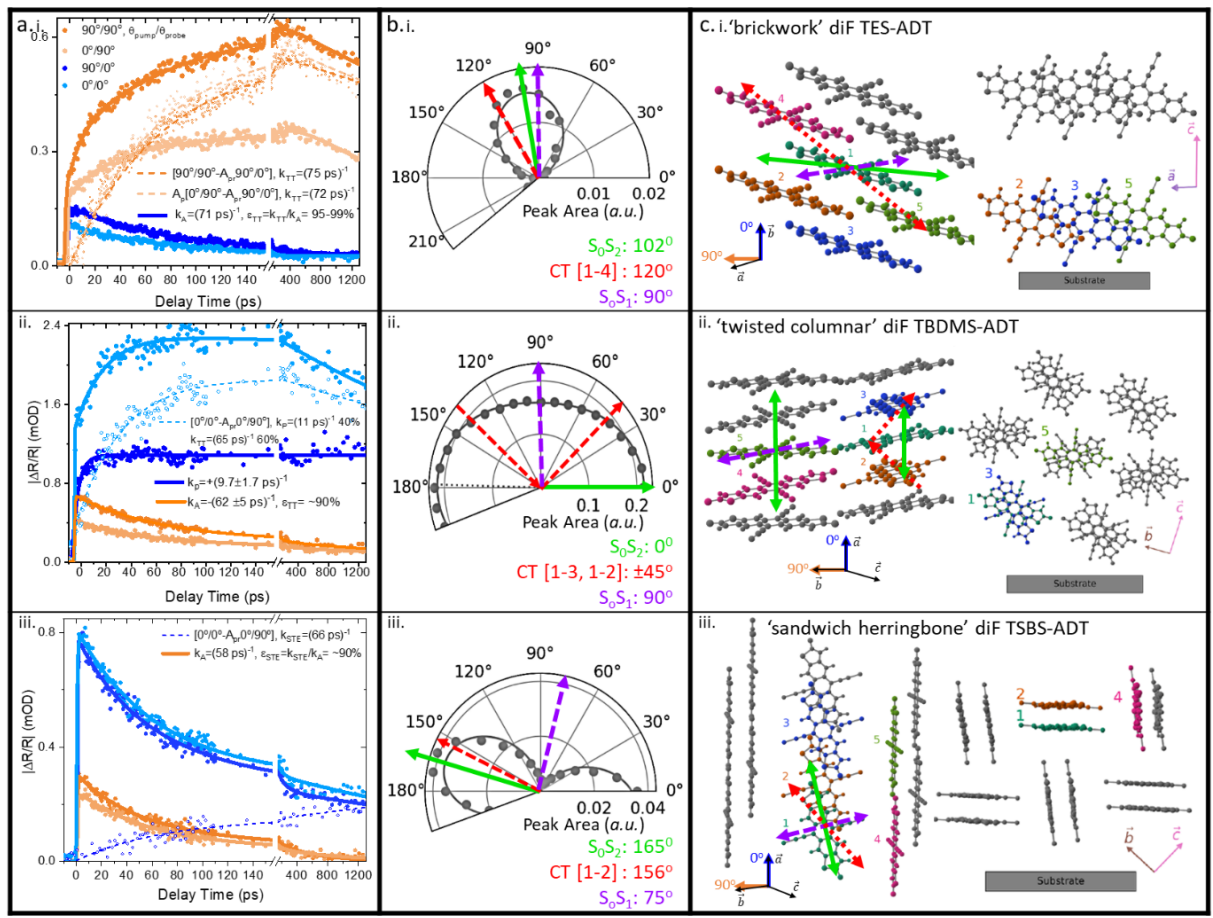

**Figure 4 - CT transition dipole axis orientation-mediated dynamics. (a)** TA relaxation dynamics of **i.** diF TES-ADT, **ii.** diF TBDMS-ADT and **iii.** diF TSBS-ADT single crystals, showing all combinations of parallel and perpendicular polarized pump and probe beams. Dashed lines are fits where the residual singlet signal is subtracted (open circles). The rates associated with these fits are in Table 3. **(b)** Corresponding polar plots of the $S_0S_2$ absorption peak integrated area are maximal (*green arrow*) for polarizations very close to the DFT calculated CT-dimer pair dipole orientation (*red dashed arrow*). *Purple dashed arrows* are in the direction of maximum pump absorption. Matching rising/falling dynamics are observed when the probe polarization is parallel to CT-axis, suggesting mixing of CT and $S_2$ states. **(c)** Top-down and side views of the diF R-ADT molecular crystals from DFT-simulations[24] with dipole orientation arrows superposed.

Further analysis of the ESA dynamics shown in Figure 4a is possible by noting that falling TA kinetics at one polarization likely still contribute somewhat to the rising kinetics taken at the opposite probe polarization. As described in the Supplementary Materials section S7, the weaker falling contribution to strongly rising kinetics may be approximately removed by curve subtraction. Figure 4a shows a dashed line that fits the falling kinetic data multiplied by a probe-beam anisotropy factor, $A_{pr}$ and subtracted from the larger rising kinetic curve. Nonetheless, fitting the resulting rises of these subtracted curves produces lifetimes in Figure 4a that are only ~3 ps longer than the *as-measured* exponential rise for diF TES-ADT. Except for diF TSBS-ADT, the rising formation kinetics overwhelm any convolved falling kinetics such that subtractive analysis shown by the dashed lines in Figure 4a is unnecessary. In Figure 4aiii, the same procedure for diF TSBS-ADT shows the 90° probe dynamics decay faster than the 0° probe data. After 66 ps, the diF TSBS-ADT electrons and lattice are fully thermalized with the surroundings, and so the residual $S_1$ relaxation is seen only for the dipole-aligned 0° probe. The ideal probe energies for each packing structure were selected using spectrally resolved TA microscopy. Specifically, the peaks of maximum ESA response were identified by taking a white-light supercontinuum probe over a 1.0 to 1.56 eV range, as shown in Figure 5b.

In summary, three of the four diF R-ADT derivatives examined gave strongly rising TA microscopy dynamics, suggesting TT formation with an average rate, $k_{TT} = (79 \pm 4 \text{ ps})^{-1}$ for a constant pump fluence of $3.5 \times 10^{12}$ photons/cm$^2$. When the probe polarization is changed by 90°, complementary falling dynamics with an average rate of $k_A=(74 \pm 6 \text{ ps})^{-1}$ are observed. Showing both the forward and reverse rates strongly contribute to $k_{TT}$, supplementary Figure S8 shows the observed diF R-ADT rise times varied from 12 ps to 174 ps as the laser fluence increased ~50x. Such stable, long-lived TT-pairs are consistent with lifetimes previously reported in other diF R-ADT systems.[15,16,24] Depending on the timescale, this TT notation may refer to coupled dipole states ($T_1T_1$) or a $T_1..T_1$ state, signifying diffusion of the triplet pair.[8,36] As this work primarily focuses on which packing morphologies facilitate vibrationally-mediated TT state formation, it does not attempt to distinguish between these various forms of triplet pair correlations.

**DISCUSSION**

I. Isolating CT-mediated singlet fission dynamics using probe polarization and energy selectivity

Isolating the rising kinetics associated with TT state formation requires careful probe energy and polarization selection. Across the visible probe energy range, diF R-ADT TA kinetics are dominated by overlapping ESA transitions, making any quantitative rate analysis challenging. While this effect is mitigated by using near-IR probes, Figure 5b shows the singlet-character of the TT state means that the TT → $S_2$ is often nearly degenerate with $S_1$ → $S_2$ ESA peaks. The TA microscopy spectra in Figure 5b plot probe energies from 1.05 to 1.35 eV and show that 1.26 eV is the optimal spectral energy to isolate the rising TT formation in diF TES-ADT.

For molecular crystals of diF R-ADT, probe polarization is also critical for optimal isolation of the TT formation kinetics. Specifically, the diffraction-limited TA microscopy measurements use probe-beam polarizations either parallel or orthogonal to the CT-dipole axis projections. The linear absorption and PLE spectra in Figures 1a, 3b and 5a, demonstrate the significance of polarization selection. Figure 5a compares the polarization-dependent absorption spectra of brickwork-packed diF TES-ADT (*top*) and sandwich herringbone-packed diF TSBS-ADT (*bottom*). In brickwork-packed diF TES-ADT, Figure 5a shows both the $S_oS_1$ and $S_oS_2$ dipole transitions are roughly maximized at 90°. By contrast, in diF TSBS-ADT the $S_oS_2$ dipole transition is maximized at 0° (*blue line*) while the $S_oS_1$ absorption peak is maximal near 90° (*orange line*). This suggests the $S_oS_2$ peak is maximized for polarizations matching the calculated CT-axis dipole projection shown in Figures 4b and c. To best observe the strongly rising TT formation kinetics in diF TES-ADT, a probe energy of 1.26 eV and 90° polarization along the short-axis of the crystal is desired.

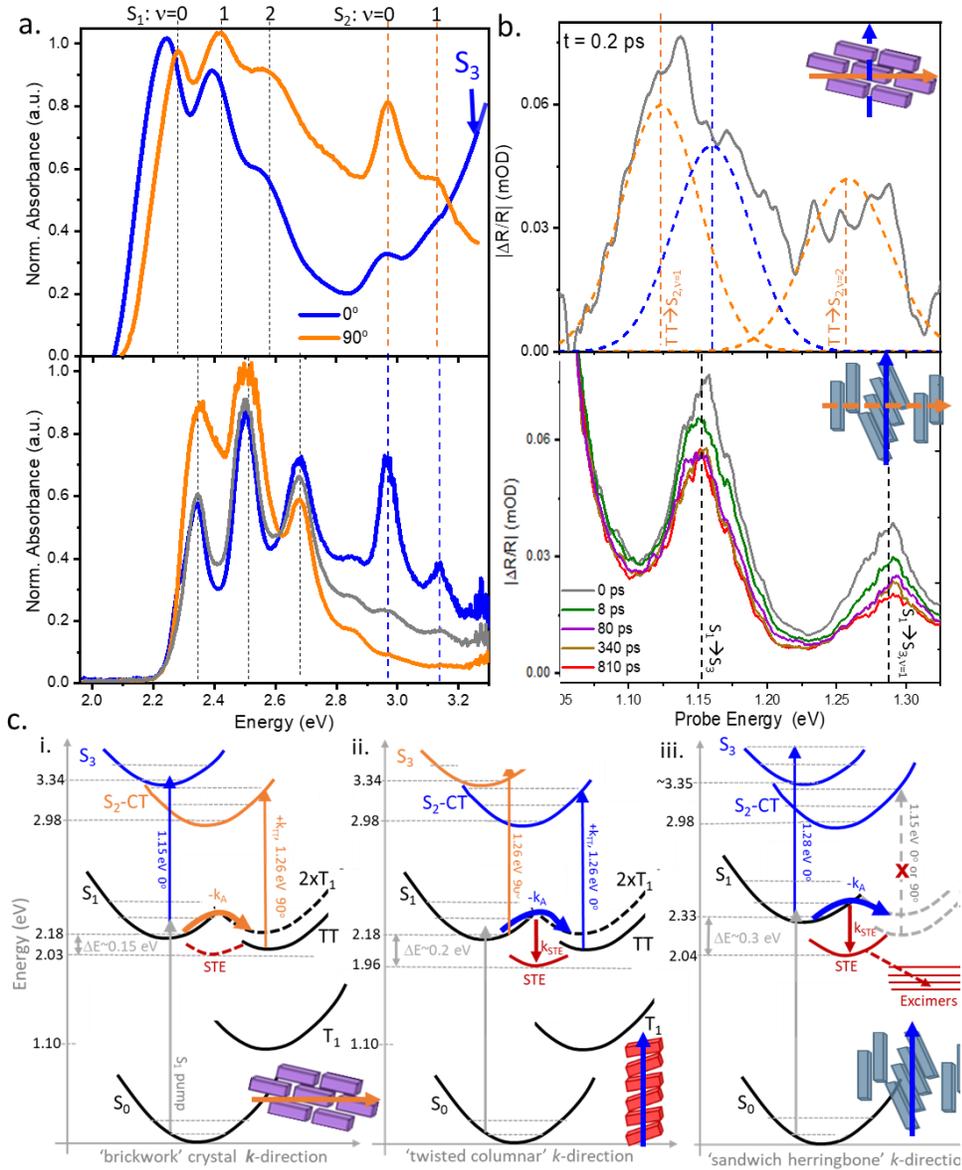

**Figure 5 - Singlet fission dynamics vs. self-trapped exciton formation. (a)** Polarization-dependent absorption spectra of diF TES-ADT (*top*) and diF TSBS-ADT (*bottom*) crystals show the CT-coupled $S_2$ absorption essentially vanishes at the polarization orthogonal to the CT-axis projection. **(b)** Single-crystal TA spectra of diF TES-ADT (*top*) and diF TSBS-ADT (*bottom*) show ESA peaks are ascribed to either $S_1$ state depletion or correlated triplet pair state (TT) formation. **(c)** Energy level diagrams highlight resonant ESA responses at $0^0$ (*blue arrow*) and $90^0$ (*orange arrow*) probe polarizations. Comparing **i.** diF TES-ADT to **iii.** diF TSBS-ADT, the latter has a larger energy barrier ($\Delta E$), making the self-trapped exciton (STE) formation pathway more favorable than TT-pair formation. **ii.** diF TBDMS-ADT is an intermediate case where TT formation is still dominant, but it completes with STE formation owing to a low-lying $CT_1$ state energy calculated.

Table 2 and Figure 1a show that while Frenkel exciton energies and Stokes shifts differ significantly between derivatives, all four derivatives have an identical $S_0S_2$ energy at 2.99 eV. This suggests that the $S_0$ and $S_2$ bands are nearly vertically aligned as the transition is strongly mixed with CT-states in all packings. The $S_0S_2$ transition drawn in Figure 5c also has a much larger polarization anisotropy factor than the $S_0S_1$ transition. Specifically, Figures 3b and 5a show the $S_0S_2$ peak disappears almost entirely whenever the polarization is orthogonal to the molecular crystal CT-axis projection. As illustrated in Figure 5ci with the TT → $S_2$ ESA transition arrow, this CT-mixing property of the $S_2$ state is critical to understanding why single-crystal diF R-ADT ESA kinetics are dominated by a rising ESA signal when resonantly probed along the CT-axis projection.

The $S_1$ depletion kinetics in all four diF R-ADT derivatives relax at a similar rate ($k_A$), which is far faster than observed in solution-phase (see Figure 1b). TA microscopy on concentrated solutions and amorphous regions generally cannot clearly isolate TT formation kinetics, as the rising TT kinetics cannot be deconvolved from competing falling $S_1$ depletion kinetics. All four derivatives share a similar energetic CT-barrier depicted in Figure 5c even if TT formation is negligible, as seen for diF TSBS-ADT. This CT-barrier is surmounted with the help of thermal and optically excited vibrational modes to ultimately generate TT pairs and self-trapped exciton (STE) states. The energetic placement of the TT state relative to the lowest CT-state helps predict the branching ratio between STE vs. TT formation using the rate conservation equation, $k_A = k_{TT} + k_{STE}$. The efficiency for TT formation over STE trapping is then calculated by the rate ratio, $\varepsilon_{TT} = k_{TT} / k_A$.

Using the energies in Table 2, Figure 5c postulates how the excited state energy landscapes may differ between each diF R-ADT packing motif. The relaxed TT state is generally accepted as lying approximately one vibronic mode below the $2E_{T1}$ energy.[24] For example, in diF TES-ADT, our DFT calculation suggests $2E_{T1}$ = 2.19 eV, and so $E_{TT} \approx 2E_{T1} - E_{vib}$, or 2.03 eV. Experimentally, this TT energy is supported by the temperature-dependent Herzberg-Teller brightened emission peaks highlighted earlier in Figure 2a and documented in other recent studies.[15,16,24,36] In Figure 5c, the CT-state energies are drawn in red to represent an electron-hole pair self-trapped between two diF R-ADT crystal sites. These CT-state values are DFT calculated for the dimer-index that gives the coupling potentials (see supplemental Table S1 for full details).[24] Table 2 below further estimates the energy barrier, $\Delta E = E_{S1} - E_{TT}$ germane to competition between TT vs. STE formation. Interestingly, $\Delta E$ increases as crystal symmetry and π-stacking character diminishes, i.e. $\Delta E$ scales with packing as R = TES < TDMS < TBDMS << TSBS. This pattern qualitatively agrees with the kinetics in Figures 3-4, showing the rising TT formation kinetic component diminishes as $\Delta E$ increases and STE formation becomes the more favorable energic pathway.

| (diF R-ADT) R= | TES (brickwork 1) | TDMS (brickwork 2) | TBDMS (twisted columnar) | TSBS (sandwich herringbone) |
|---|---|---|---|---|
| $S_0S_1$ Abs (eV) | 2.25 | 2.26 | 2.25 | 2.36 |
| $S_0S_2$ Abs (eV) | 2.98 | 2.98 | 2.98 | 2.98 |
| Stokes shift (eV) | 0.13 | 0.13 | 0.14 | 0.05 |
| $S_1$ Frenkel (eV) | 2.19 | 2.20 | 2.18 | 2.33 |
| $E_{TT}$ TT pair (eV) | 2.03 | 2.04 | 2.04 | 2.19 |
| $E_{CT}$ (eV) [dimer in] | 2.032 [1-2] | - | 1.954 [1-3] | 2.04 [2-3] |
| $\Delta E = E_{S1} - E_{CT}$ (eV) | 0.16 | - | 0.22 | 0.29 |
| $E_{TT} - E_{CT}$ (eV) | 0.02 | - | 0.07 | 0.16 |

**Table 2** The energy states above are also drawn in Figure 5c, and $\Delta E$ above helps predict which crystal packings self-trap $S_1$ Frenkel excitons. Values are experimentally determined except for CT energies, which were DFT calculated for the optimal crystal dimer index labeled.[24]

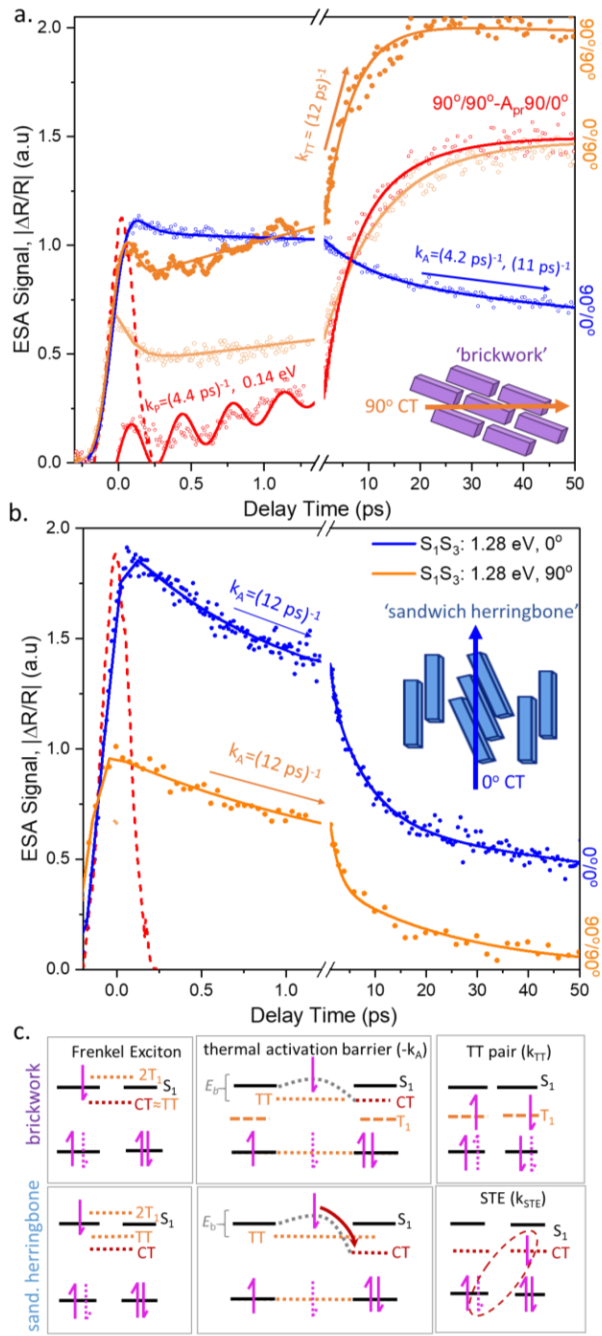

**Figure 6 - Ultrafast thermally and vibrationally CT-mediated dynamics. (a)** TA microscopy of diF TES-ADT (1.26 eV probe) at $\theta_{pump}/\theta_{probe}$ polarizations labeled. When pumped at 0/90°, the signal rises with a 12.3 ps time constant; however, when pumped along the CT-axis at 90°/90° the rise is instead biexponential. Weak vibrational beating at 33 ps period is best seen in the subtracted data [90°/90° – $A_{pr}$ 90°/0°] fitted in red. **(b)** diF TSBS-ADT dynamics (1.28 eV resonant probe) fall at similar rates $k_A$, but there are no rising kinetic contributions at any probe energy (1.15 to 1.38 eV) or polarization. **(c)** (*top*) Schematic of endoergic singlet fission in brickwork-packed diF TES-ADT. The equivalent STE process observed for diF TSBS-ADT packing, (*bottom*) shows TT formation kinetics become unfavorable when the $S_1$ to the lowest-lying CT state separation is much larger than the $S_1$-TT energy separation.

## II. Thermally activated kinetics: self-trapped exciton vs. TT formation

While anthradithiophene is best classified as an endoergic singlet fission material, the energetic barrier between $S_1$ and $2 \times T_1$ can be surmounted by either thermal activation or faster vibrationally-mediated routes. Figures 6a-b compare the ultrafast TA relaxation kinetics of diF TES-ADT to diF TSBS-ADT crystals at 0° (*blue circles*) and 90° (*orange circles*) probe polarizations. The instrument response (*dashed line*) results from a convolved resonant optical parametric amplifier (OPA) pump at 2.35 eV, and a TT → $S_2$ resonant non-collinear optical parametric amplifier (NOPA) probe at $E_{pr}$ = 1.26 eV. Both Figures 6a-b show dynamics on the order of 12 ps that are faster than the ~76 ps timescales reported in Table 1 because the pump photon flux is ~15x larger at $9.4 \times 10^{13}$ photons/cm$^2$. This rate acceleration is induced by laser heating and the net reverse rate processes (e.g. singlet fusion and triplet-triplet annihilation), which have been previously observed in diF R-ADT materials.[15]

Over the 0-1.2 ps interval plotted in Figure 6a (before break), only the 90° (orange circles), CT-axis-probed data shows a fast initial decay spike, $k_{fast}$= (0.102 ps)$^{-1}$. This fast decay is largely absent at 0° probe polarization (*blue circles*). As the observation of this fast initial spike requires selective probe-beam polarization along the CT-axis, we connect it with fast TT formation processes before the Frenkel exciton state has thermalized. Similar fast decays are prominent in other endoergic singlet fission materials such as tetracene.[37] Consistent with singlet fission being prohibited in the sandwich herringbone-packing, Figure 6b shows this $k_{fast}$ component is also absent for the diF TSBS-ADT crystal (*blue circles*).

On the picosecond timescale, the 90° pump polarization kinetics (*dark orange*) in Figure 6a rise faster than the 0° pump polarization kinetics (*light orange*). A monoexponential initial rise and fall is needed to fit the 0° pump case, giving $k_{TT}$= (12 ps)$^{-1}$ and $k_A$= (11 ps)$^{-1}$. When pumped at 90° along the CT-axis, the rising and falling kinetics of diF TES-ADT in Figure 6a require a biexponential fit with an additional 4.8 ps rise component that comprises ~15% of the overall rising amplitude. The introduction of a second rising rate, $k_p$= (4.8 ps)$^{-1}$ is further supported by the weak beating fitted in the 90° pump TA microscopy kinetics plotted in Figure 6a (*red trace*). While our 72 fs instrument response could not fully resolve the ~33 fs phonon beating mode shown, this beat period corresponds with the $E_{vib}$ = 0.14 eV C-C vibronic mode of diF-R ADT. As the mode is observed in the rising TT formation kinetics, it suggests that optically excited vibrational modes are mediating TT formation at faster rate, $k_P$. Interestingly, when the pump is orthogonal to the CT-axis projection of diF TES-ADT, this faster rise is no longer observed (see 0° pump, 90° probe kinetics), suggesting that these coherent vibrational modes only help mediate TT formation along the CT-axis.

Comparing the TA microscopy spectral peak widths shown in Figure 5b, the diF TES-ADT ESA peaks are ~1.8x broader than the diF TSBS-ADT ESA peaks. The dashed gaussian lines in Figure 5b suggest diF TES-ADT has $S_1 \rightarrow S_n$ transitions that are convolved with the $TT \rightarrow S_2$ ESA peaks. The diF TSBS-ADT packing ESA peak widths are narrower as TT state formation is not expected. The resulting kinetic trace probed at the $E_{pr}$ = ~1.15 eV ESA resonance is shown in Figure 7b. Here, the relaxation at 90° probe polarization is slower than at 0° because the falling singlet and rising triplet pair kinetics strongly overlap at the $E_{pr}$ = ~1.15 eV ESA peak in Figure 5b. Upon subtracting the two curves, the red-trace in Figure 7b shows the rising dynamics successfully recover the same 74 ps rise originally observed at the optimal 1.26 eV probe energy. Contributions from singlet depletion kinetics are minimal near the $E_{pr}$ = 1.26 eV ESA peak, as it requires an unlikely $S_1$ to $S_2$ (v = 3) ESA transition. Thus, the strongly rising triplet pair ESA dynamics are best resolved at the 1.26 eV probe energy with probe polarization along the CT-axis projection of either brickwork or twisted columnar packing motifs.

Compared to dilute solution measurements with negligible CT-mediated processes, the $S_1$ state populations ($n_{S1}$) shown in Figures 6a-b are depleted rapidly by vibrationally-assisted ($k_p$) and thermally-activated rates, $k_A$. The depletion of the $S_1$ state on intermediate timescales of ~0.5 ps (post-thermalization) to several nanoseconds is approximated by the rate equation

$\frac{dn_{S1}}{dt} = \delta(t)n_o - (k_{TT} + k_{STE})n_{S1}$, where $n_o$ is the thermalized singlet initial population. At higher photon flux, the contribution from optical vibrational modes becomes increasingly significant, and a second faster rate contribution, namely $k_P = k_{TT}' + k_{STE}'$, results in a biexponential rise. The contribution of this faster singlet depletion rate process $k_p$, is small enough to be neglected with its rising amplitude not exceeding 15%. On a longer nanosecond timescale, the radiative and non-radiative rates, $k_{S1} = k_r + k_{nr}$, deplete the $S_1$ Frenkel exciton state. These dynamics are studied extensively elsewhere.[23]

Since strongly rising kinetics in Figures 3-6 suggests the probe beam predominantly resolves TT-state formation, a different kinetic model is required when the probe beam is polarized along the CT molecular crystal-axis projection. The triplet pair state population ($n_{TT}$) is approximated by the $S_1$ coupled rate law, $\frac{dn_{TT}}{dt} = k_{TT}n_{S1} - (k_{HT} + k_{T1})n_{TT}$. $k_{HT}$ refers to the radiative relaxation of the TT state, predicted in diF R-ADT to occur by Herzberg-Teller PL emission.[15,16,24,36] Full details and solutions to the coupled kinetic rate laws are found in Supplemental Materials section S6. Applying this rate model of CT-mediated processes to all four derivatives gives the rates summarized in Table 3 below. The rate ratios also give the TT-formation efficiency, $\varepsilon_{TT}$ relative to the competing self-trapped exciton (STE) pathways.

| (diF R-ADT) R= | TES (brickwork 1) | TDMS (brickwork 2) | TBDMS twisted columnar | TSBS (sandwich herringbone) | Description |
|---|---|---|---|---|---|
| $\theta_{pump}$, $S_oS_1$ | 90° | 90° | 90° | 75° | $S_oS_1$ max pol. |
| $\theta_{pr}$, $S_1S_2$ | 90° | 90° | 0° | 0° | CT-axis max pol. |
| $k_P\uparrow$ps$^{-1}$ (% rise) | Not needed | Not needed | 0.11 (~40%) | N/A | TT rise, op. phonon |
| $k_A\downarrow$ns$^{-1}$ [$\theta_{pr}$] | 13±0.3 [0°] | 13±1.5 [0°] | 16.1±1.4 [90°] | 16±0.5 [0°, 90°] | $S_1$ fall, thermal ac. |
| $k_{TT}\uparrow$ns$^{-1}$ [$\theta_{pr}$] | 13±0.6 [90°] | 11±0.9 [90°] | 15.4±1.6 [0°] | N/A | TT rise, thermal ac. |
| $\varepsilon_{TT} = k_{TT}/k_A$ | ~95-100% | ~84% | ~88% | ~0% or N/A | TT vs. STE efficiency |
| $k_{HT}\downarrow$ns$^{-1}$ [$\theta_{pr}$] | 0.14±0.02 [90°] | 0.11±0.02 [90°] | 0.38±0.05 [0°] | N/A | TT fast decay |
| $k_{S1}\downarrow$ns$^{-1}$ [$\theta_{pr}$] | 0.32±0.10 [0°] | 0.29±0.10 [0°] | 0.33±0.03 [90°] | 0.29±0.05[0°,90°] | $S_1S_o$ decay, $k_r+k_{nr}$ |
| $k_{HT}+k_{S1}(k_{PL})$ns$^{-1}$ | 0.46±0.04(0.48) | 0.40±0.04 (-) | 0.71±0.05 (0.6) | 0.29±0.05(0.37) | comp to PL rates[24] |

**Table 3** Summary of the rising($\uparrow$) and falling ($\downarrow$) packing morphology-dependent rates extracted from all data shown in Figures 3-4 (see Supplement Section S6 for the full kinetic model). The last row shows our longer TA rates are similar to recently reported TCSPC-measured PL rates ($k_{PL}$ is $k_2$ from Ref. 24). All measurements were taken at room temperature under high-vacuum.

III. Impact of diF R-ADT packing morphology on singlet fission efficiency

As diagramed in Figure 6c, endoergic triplet pair formation can proceed by a concerted spin-conserved superexchange[39,40] of electrons (*solid*) and holes (*dashed*) to generate the final triplet pair state drawn. In brickwork and twisted columnar packings, the lowest CT-state and the relaxed TT states are approximately within $k_bT_l$ of each other, making the concerted exchange of charges to form TT pairs the most favorable scattering route. The opposite is true for the sandwich herringbone CT-mediated processes that preferentially trap carriers.

Interestingly, Table 2 shows the calculated CT states of diF TES-ADT and diF TSBS-ADT packing are approximately degenerate, and so the higher Frenkel exciton energy of the sandwich herringbone structure becomes a critical factor. Specifically, Figure 6c summarizes how diF TSBS-ADT preferentially scatters to the CT-state calculated to be $\Delta E$ ~110 meV lower than the TT state. Large $\Delta E$ values should be more common in crystals and crystallites with mixed packing symmetries like the sandwich herringbone molecular packing, and serve to preferentially trap carriers that would otherwise proceed to form stable, long-lived triplet pairs.

Table 3 above summarizes the rate constants obtained for the various packing morphologies and lists the probe polarization conditions that maximize (or minimize) the rising or falling kinetic contributions. The rate constants associated with TT state were separated from the $S_1$ decay rates by simply rotating the probe-beam half-waveplate 90°. The longer rate constants ($k_{S1}$, $k_{HT}$) in Table 3 are further compared with recently published room-temperature rates ($k_{PL1,2,3}$) obtained from time-resolved single-photon counting (TCSPC) photoluminescence rates.[24] Specifically, we observe rate equivalence between the two methods if $k_{PL2} \approx k_{S1} + k_{HT}$.[24] Unlike the TA microscopy rates, the TCSPC-measured rates cannot be separated based on polarization, and so the second TCSPC rate $k_{PL2}$ reflects the sum of TT-relaxation and $S_1$-relaxation rates ($k_{HT} + k_{S1}$) obtained by polarization-dependent TA microscopy.

The rates in Table 3 were all taken at a low pump fluence of $3.5 \times 10^{12}$ photon/cm², and are approximated well by a monoexponential rise and fall. Only the kinetics of diF TBDMS-ADT had a % amplitude large enough (at ~40%), that both $k_{TT}$ and $k_p$ rates were necessary. As the rates associated with thermally-activated singlet fission accelerate with pump fluence, the rate ratio or efficiency ($\varepsilon_{TT} = k_{TT} / k_A$) provides a more meaningful quantitative metric. $\varepsilon_{TT}$ is the efficiency of TT state formation relative to STE formation. Table 3 suggests the TT formation process in the diF TES-ADT brickwork packing is the most efficient at 95-100%. Compared to diF TES-ADT, the R- substituent side groups in Figure 1a shows diF TDMS-ADT contains three extra –(CH$_2$)CH$_3$ groups. This suggests that the bulkier TDMS substituent makes self-trapping more favorable, lowering $\varepsilon_{TT}$ to 84% as summarized in Figure 7a.

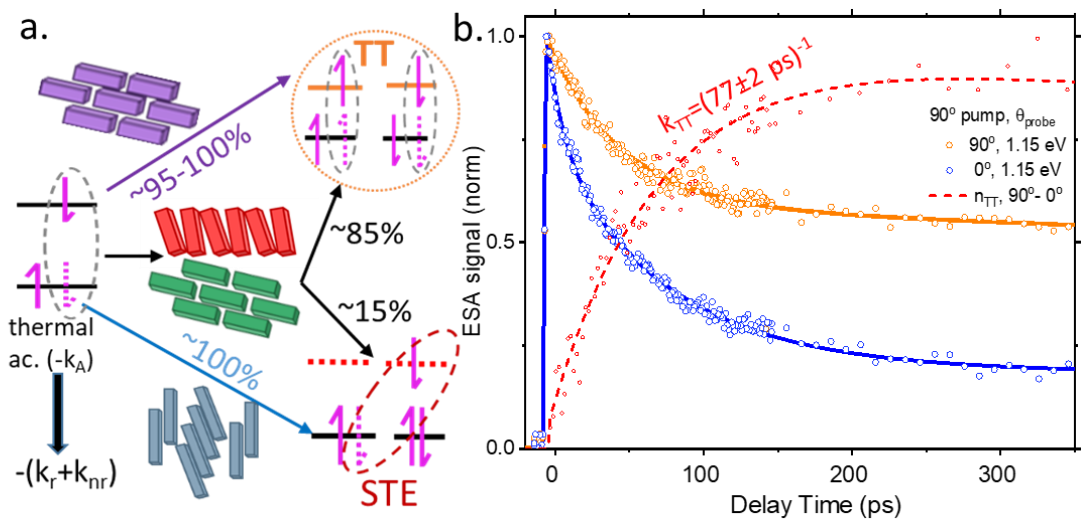

**Figure 7 - Packing morphology-dependent TT-pair formation efficiency**, $\varepsilon_{TT} = k_{TT} / k_A$. **(a)** Comparison of the packing-morphology dependent branching ratio for TT formation with self-trapped exciton formation (STE) estimated by taking the ratios of the rising and falling TA rates. **(b)** TA microscopy of diF TES-ADT with $E_{pr}$ = 1.15 eV contains convolved dynamics that are separated by subtractive analysis [90°/90° – $A_{pr}$ 90°/0°, shown in red] to recover the 77 ps rise seen directly for $E_{pr}$ = 1.26 eV in Figures 3-4.

Despite its favorable π-stacking configuration, with an $\varepsilon_{TT}$ of 88%, twisted columnar-packed diF TBDMS-ADT is more likely to trap excitons than the brickwork-packed diF TES-ADT. This is likely owing to its lower-lying $CT_1$ state energy shown in Figure 5cii and Table 2. As one caveat, diF TBDMS-ADT consistently gave the fastest kinetic rises owing to enhanced contribution from vibrationally-mediated TT formation with rate $k_p$. This $k_p$ rate process is observed most strongly when both the exciton and vibrational mode are excited along the CT-axis (0° pump polarization). This suggests that vibrationally coherent exciton-phonon processes may mediate TT formation better in the twisted columnar packing than in brickwork.

No clear evidence of TT state formation could be resolved in diF TSBS-ADT crystals at any energy or polarization. The subtractive analysis shown in Figure 4c suggests that if the contribution exists at all, it must be very small in amplitude compared to the self-trapping efficiency implied by the falling singlet dynamics. As one last consideration, the CT character of a material, or how much the CT and Frenkel excitons mix, can enhance singlet fission.[41,42]. diF TSBS-ADT retains the most Frenkel character, making it unfavorable for TT formation. diF TSBS-ADT has only two nearest neighbors with a large slip distance between adjacent molecules in the "sandwich" part of the "sandwich-herringbone" packing. DFT predicts that diF TSBS-ADT has the smallest total exciton coupling of the three systems. As a result, in contrast to the other derivatives, the exciton delocalization in diF TSBS-ADT is suppressed, and so rather than forming a TT state, the formation of highly localized Frenkel exciton and stable excimer are favored instead.[24] In diF TES-ADT, large exciton couplings that lead to Frenkel-CT mixing have been attributed to its 2D nature which provides more nearest neighbor molecules. diF TBDMS-ADT also has a pronounced CT character, making it favorable to undergo singlet fission. However, rather than the number of nearest neighbors, its large π-π overlap between adjacent molecules that results in a smaller slip distance between molecules in the column is what leads to high interactions. This may help explain why the optically-driven phonon rate, $k_p$ that mediates TT formation was ~8x stronger in amplitude than seen in the brickwork packing.

In summary, unlike brickwork or twisted-columnar packing motifs, the sandwich herringbone packing shows little evidence of either singlet fission or TT formation. This is supported by: (i) the absence of any ultrafast $k_{fast}$ dynamic rates in diF TSBS-ADT kinetics, suggesting any direct singlet fission pathways are inhibited by more localized excitations, (ii) the $S_1$ to lowest $CT_1$ state relaxation energy, $\Delta E$ is ~2x larger in diF TSBS-ADT, suppressing any CT-mediated TT state formation, (iii) brickwork packing geometries have both multi-directional translational symmetry in the incident plane of polarization and more nearest-neighbors than sandwich herringbone packing (four in diF TES-ADT vs. two in diF TSBS-ADT).

**CONCLUSIONS**

While diF R-ADT is best classified as an endoergic singlet fission material, the energetic barrier between $S_1$ and $2\times T_1$ states is sufficiently small that both vibrational and thermally-activated population of the TT-pair state is possible.[36,42] The TT-pair formation rate is limited by an $S_1$-CT energy barrier that depletes the singlet state with a thermally activated net bidirectional rate, $k_A$. Using polarization-dependent TA microscopy, the corresponding rate of TT state formation, $k_{TT}$ was isolated in three of the four single-crystal functionalized derivatives of diF R-ADT (R = TES, TDMS, TBDMS, TSBS). Certain packing morphologies preferentially form self-trap excitons (STE) in excimer-like states at a rate calculated by $k_{STE} = k_A - k_{TT}$. For each molecular packing, the competing rising rate, $k_{TT}$ and falling rate, $k_A$ were separated using probe beam polarizations that are either parallel or orthogonal to the single-crystal CT-axis projection. As was shown in Figure 5b, the resonant probe window further was carefully selected to find a TT→$S_2$ ESA peak that was not convolved with $S_1$→$S_n$ peaks. By combining careful probe-energy and probe-polarization selectivity, TA microscopy successfully isolated falling singlet and rising TT dynamics from the raw TA traces, without any need for error-prone post-analysis (e.g. by subtracting kinetics curves, or pump-probe anisotropy). The ratio between rising and falling rates $\varepsilon_{TT} = k_{TT} / k_A = k_{TT} / (k_{TT} + k_{STE})$ is equivalent to the efficiency of TT formation over CT-mediated self-trapped exciton formation.

In three of the four molecular crystals studied, the TT formation efficiency was large ($\varepsilon_{TT}$ = 84% to nearly 100 %). Brickwork and twisted-columnar packings give efficiencies that decrease with the energy separation $\Delta E = E_{S1} - E_{CT}$ and the relative size of the -R group substituent. diF TES-ADT and diF TDMS-ADT both exhibit 2D brickwork packing structures and are dominated by a strongly rising TA response that accelerates from 153 ps to 12 ps with increasing photon flux. Likewise, when resonantly probed parallel to the charge transfer (CT)-axis, similar rising timescales were extracted with the 1D twisted-columnar packing structure of diF TBDMS-ADT. The probe polarization-dependent dynamics suggest that the excited-state TT transition dipole orientation is perpendicular to the long axis of the diF TES-ADT and diF TDMS-ADT crystals and the short axis of the diF TBDMS-ADT crystal. At higher pump photon fluxes, the net TT formation could not be described by a single rate and required a biexponential rising fit. This second fast rate, $k_p$ is associated with coherent vibrationally-assisted TT formation and was shown in Figure 6 to be most prominent when the vibrationally excited mode was pumped parallel to the crystal CT-axis. In diF TES-ADT, the $k_p$ rate contribution to TT formation was always small (<~15% of rising amplitude). Interestingly, in diF TBDMS-ADT, this coherent vibrational contribution to overall rising kinetics was at least 2x larger.

Lastly, for diF R-ADT molecules with a sandwich-herringbone motif, the dynamic signatures of singlet fission are almost completely suppressed. The geometry of diF TSBS-ADT leads to more localized excitons, which inhibits a coherent formation of the TT state. It also leads to a greater energetic barrier between $S_1$ and TT, favoring other relaxation channels. These attributes together suggest that the sandwich-herringbone packing motif is prohibitive for endoergic singlet fission. While diF TES-ADT remains a promising singlet fission organic semiconductor, we have shown that small changes to the crystal packing result in preferential exciton self-trapping rather than charge multiplication. Thus the barrier for TT-formation is ordinately sensitive to subtle tweaks in the molecular packing. This motivates futher work exploring how external temperature and applied fields may further maximize both TT-pair formation and triplet extraction yields for diF R-ADT derivative and related endoergic singlet fission systems.

**Acknowledgments**

This work was in part supported by the National Science Foundation (Grant No. and DMR-1808258 and DMR-1920368). We gratefully acknowledge Mirek Brandt for developing the hyperspectral photoluminescence single-crystal excitation spectra measurements.

# Supplementary Information:
# Anthradithiophene crystals with high translation symmetry promote thermally-activated singlet fission

Gina Mayonado[1], Kyle T. Vogt[1], Jonathan D. B. Van Schenck[1], Liangdong Zhu[1], John Anthony[2], Oksana Ostroverkhova[1], Matt W. Graham[1]

1-Department of Physics, Oregon State University, Corvallis, Oregon 97330, USA
2-Department of Chemistry, University of Kentucky, Lexington, Kentucky 40506, USA

## S.1 Solution Measurements

Figure S1 shows TA measurements of dichlorobenzene solutions of different diF R-ADT derivatives. All traces show similar decay kinetic behavior with a ~600 ps decay time for an ~0.1 mM concentration. Therefore, the functionalization of the diF-ADT backbone does not significantly alter the monomer decay dynamics.

The decay in Fig. S1 was invariant to probe wavelength for the probe wavelengths measured (between 935nm and 1010nm). In more concentrated diF TES-ADT solutions, this decay process still dominates but at a faster rate. The rates of singlet fission are still much lower than in the crystal phase due to the disorder but appear to occur on similar timescales.

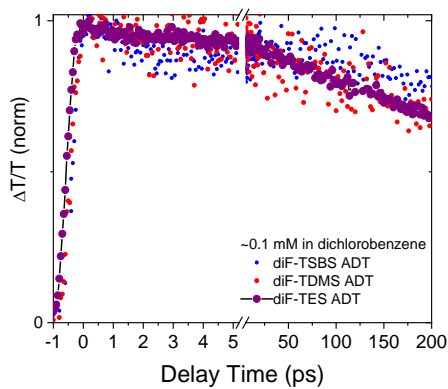

**Figure S1:** TA kinetic decay kinetics of ~0.1 mM dichlorobenzene solutions of diF TES-ADT (*purple*), diF TDMS-ADT (*red*,) and diF TSBS-ADT (*blue*) are approximately invariant with an ~600 ps decay rate.

## S.2 X-Ray Diffraction

X-ray diffraction measurements were taken for diF R-ADT crystals to confirm the presence of regions with sufficiently high crystallinity for diffraction-limited single-domain measurements.

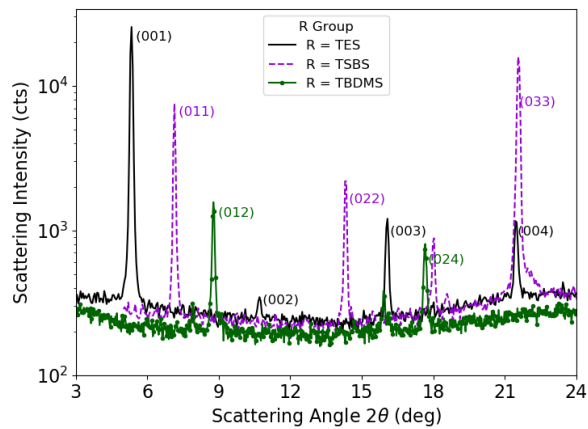

**Figure S2:** X-ray diffraction measured for diF R-ADT (R = TES, TSBS, TBDMS) crystals grown on a glass substrate. The well-defined peak structures confirm the crystal nature of the materials, and the corresponding Miller indices for each peak are labeled.[1]

## S.3 Crystal Boundaries

diF R-ADT crystals were first analyzed by polarization-dependent hyperspectral imaging microscopy to (i) first identify nearly single-crystal domains as shown in Fig. S3, and (ii) obtain polarization-dependent absorption and photoluminescence excitation spectra.

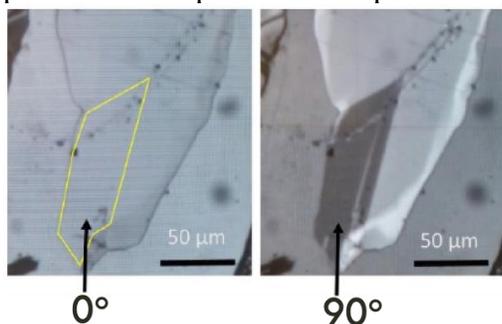

**Figure S3:** Clear diF TES-ADT crystal boundaries visible using polarization-dependent microscopy for 0° and 90° polarizations.

Transient absorption microscopy was utilized to focus the beams within a single crystalline domain which were identified by their different polarization-dependent optical responses. Measuring within single crystal boundaries was critical to isolating individual physical processes, rather than averaging over many domains which may have different orientations and obscure the dynamics. **Please see *Supplementary Video 1*,** for ultrafast microscopy movies taken over 800 ps that show diF TES-ADT exhibits spatially uniformly rising and falling ESA responses that depend on the selected probe polarization of 0° and 90°.

## S.4 Transient Absorption Microscopy Experimental Setup

Transient absorption microscopy measurements of diF R-ADT crystals were taken using the setups illustrated below. Details are included in the main text.

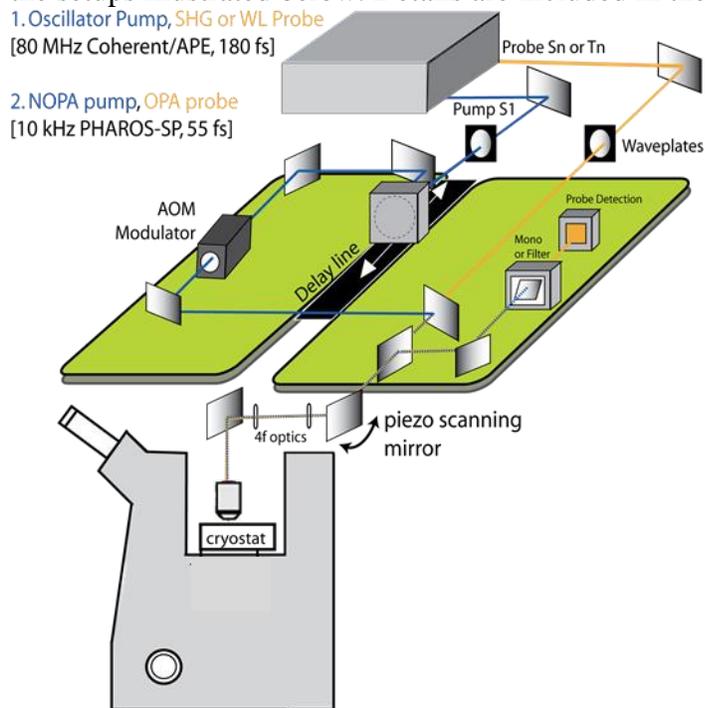

**Figure S5:** Transient absorption microscopy measurements were taken using the pump-probe setup illustrated above, utilizing either: (1) ~180 fs pulses at 80 MHz from a second-harmonic generation system (SHG, APE HarmoniXX) pumped by a Coherent Ultra Chameleon, or (2) a 35 to 95 fs pulses at 10 kHz from a non-collinear optical parametric amplifier (NOPA, ORPHEUS-N, LightConversion), raster-scanned through either (1) a 50XIR objective or (2) a 15X reflective objective, and the transient reflection signal retrieved using lock-in detection.

## S.5 diF R-ADT Charge Transfer State and ESA Energies

Table S1 below shows various transition energies associated with diF TSBS-ADT, diF TES-ADT, diF TBDMS-ADT and diF TDMS-ADT. These values along with table 2.0 were used to illustrate the energy level diagrams in Fig. 5c. The CT-state energies were DFT calculated for specific monomer pairs. Here, we present only the CT states with sufficiently large coupling, electron/hole transfer integrals ($t_e$ / $t_h$), and oscillator strengths below. More details are available in the Supplemental section of Schenck et al.[1]

| *Side Group, R=* | **TSBS** *(s. herringbone)* | **TES** *(brickwork)* | **TBDMS** *(t. columnar)* | **TDMS** *(brickwork)* |
|---|---|---|---|---|
| *Stokes Shift, $E_S$ (eV)* | 0.058 | 0.13 | 0.078 | 0.13 |
| *$S_1$ Frenkel energy (eV)* [$E_{S1}$- ½ $E_S$] | 2.33 | 2.19 | 2.18 | 2.20 |
| *TT energy (eV) relaxed* [$2xT_1$- $E_{vib}$] | ~2.20 | 2.03 | 2.04 | 2.04 |
| *CT energy* [dimer index] | 2.043 [2-3]<br>2.0725 [1-2] | 2.032 [1-4]<br>2.064 [1-2] | 1.954 [1-3] | -- |
| *CT oscillator avg. strength* | 0.0313 | 0.0338 | 0.0135 | -- |
| *Coupling strength |V| (meV)* | 4 [2-3], 2 [1-2] | 7 [2-3], 20 [1-2] | 52 [1-3] | -- |
| *|$t_e$|/|$t_h$| (meV)* [dimer index] | 41/ 27 [2-3]<br>23/11 [1-2] | 2 / 68 [1-4]<br><1 / <1 [1-3] | 63 / 20 [1-3] | -- |
| *TT→$S_2$ v=0 ESA* | 0.98 | 1.00 | 1.06 | 1.00 |
| *TT→$S_2$ v=1 ESA* | 1.12 | 1.14 | 1.20 | 1.14 |
| *TT→$S_2$ v=2 ESA* | 1.26 | 1.27 | 1.34 | 1.27 |
| *$S_1$→$S_2$ v=0 ESA* | 0.71 | 0.86 | 0.81 | 0.86 |
| *$S_1$→$S_2$ v=1 ESA* | 0.85 | 1.00 | 0.95 | 1.00 |
| *$S_1$→$S_2$ v=2 ESA* | 0.99 | 1.14 | 1.09 | 1.14 |

**Table S1**: Summary of the energies associated with ESA pump-probe responses for each packing morphology of ADT. Full detail of DFT calculations of CT dimer state calculations can be found in Schenck et al.[1]

## S.6 Kinetic Model Details

When any fast sub-picosecond dynamics ($k_{fast}$) seen in Fig. 6a are ignored, only two exponents in a least-squares deconvolution fit to TA kinetics plotted in Fig. 3 are needed to minimize residuals. Specifically, when resonantly probed along the CT-axis projection, to acceptably minimize the fit residuals, we require:
(i.) a single exponential rising and falling component for brickwork packings,
(ii.) a biexponential rising and monoexponential falling component for twisted columnar packing
(iii.) a biexponential falling fit for sandwich-herringbone packing or diF-ADT TSBS.

At very high photon fluxes such as in Fig. 6a-b, the faster optical phonon-assisted TT formation rates ($k_p$) become significant (up to 20% of the rising amplitude). Otherwise, the inclusion of more rate components did not improve the fit residuals. Motivated by these empirical observations, the global kinetic fit model approximation is outlined in Fig. S5 and Table S2 below.

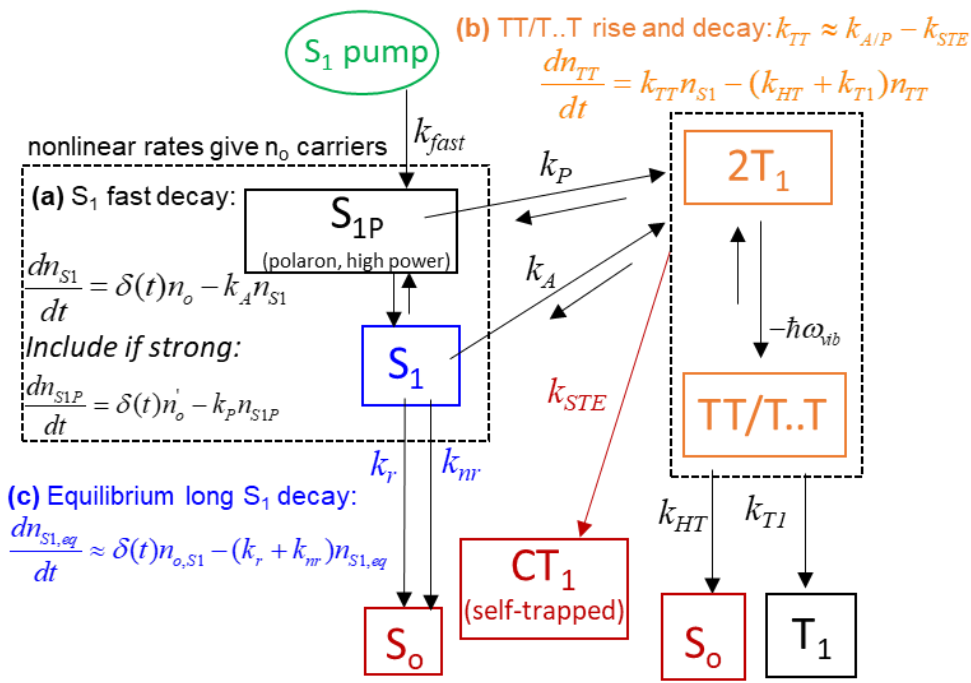

**Figure S6**: Schematic of kinetic rate model employed, yield two sets of observable rate laws that coincide with the TT state (orange) and $S_1$ state population (blue). Probe polarization is used to distinguish the TT dynamics from $S_1$ kinetics. The photon flux dependence is not explicitly included in this model, we instead assume every incident pump power establishes an equilibrium rate for the forward and reverse kinetic processes (see bidirectional arrows).

The below table accompanies Fig. S6 above to show how processes labeled a, b, and c are evaluated and elucidate under what conditions they are observed in data. Sub-picosecond rates ($k_{fast}$) before the exciton population thermalizes are ignored, but are discussed in the main text and fitted in Fig. 6a.

|   | Rate Equations | Carrier Density Dynamics | Conditions for Observing |
|---|---|---|---|
| (a) | i. $\frac{dn_{S1}}{dt} = \delta(t)n_o - k_A n_{S1}$ (thermal ac. rate. $k_A$) ii. $\frac{dn_{S1P}}{dt} = \delta(t)n_o' - k_P n_{S1P}$ (optical phonon rate, $k_P$) | i. $n_{S1}(t) = n_o e^{-k_A t}$ ii. $n_{S1P}(t) = n_o' e^{-k_P t}$ (optically-excited phonon-assisted rate $k_P$ is only observed at higher pump powers and in diF-ADT TBDMS) | Present in all data. Dominates signal if probe energy or polarization is non-resonant. The optically-excited phonon-assisted rate, $k_P$ can generally be ignored for low pump laser fluxes. (see (d)*). |
| (b) | $\frac{dn_{TT}}{dt} = k_{TT} n_{S1} - (k_{HT} + k_{T1}) n_{TT}$ $k_{TT} \approx k_A - k_{STE}$ | $n_{TT}(t) = n_o \left( \frac{k_{TT}}{k_{TT} - (k_{HT} + k_{T1})} \right) \left( e^{-(k_{HT}+k_{T1})t} - e^{-k_{TT}t} \right)$ (TT rise and decay) | Dominant when probe beam is polarized parallel to CT-axis and $S_1 S_2$ resonant if TT states form |
| (c) | $\frac{dn_{S1,eq}}{dt} \approx \delta(t)n_{o,S1} - (k_r + k_{nr}) n_{S1,eq}$ (eq. long $S_1$ decay) | $n_{S1,eq}(t) \approx n_{o,S1} e^{-(k_r+k_{nr})t}$ | Dominant when probe beam is polarized *anti*-parallel to CT-axis projection and $S_1 S_3$ resonant |
| (d)* | $\frac{dn_{TT}}{dt} = (k_P - k_{STE}) n_{S1P} - k_{TT} n_{S1} - (k_{HT} + k_{T1}) n_{TT}$ | $n_{TT}(t) = -A_1 e^{-(k_P-k_{STE})t} - A_2 e^{-k_{TT}t} + A_3 e^{-(k_{STE}+k_{HT})t}$ general triexponential approximation | Some diF-ADT packing (esp. π-stacked, twisted columnar) have contributions from rate $k_P$. In such cases, a generalized triexponential fit is used instead. |

**Table S2:** Table of rate equations, carrier density dynamics, and conditions used for the kinetic rate model applied to data set of varying probe polarization.

When the crystal is resonantly probed along the CT-axis projection, the kinetics are dominated by the rising triplet pair formation rate $k_{rise} = k_{TT}$, and the decay of the triplet pair $k_{fall} = k_{HT} + k_{T1}$. When the crystal is resonantly probed perpendicular to the CT-axis projection, we are instead predominately sensitive to $S_1$ state depletion at rate $k_{fall1} = k_A$ for the CT-mediated thermally activated depletion rate, and rate $k_{fall2} = k_r + k_{nr}$ for monomer radiative and non-radiative relaxation. The self-trapped excitons (STE) population is not directly observed as the STE-states are expected to be nearly optically dark to the ESA probe applied. However, the resulting exciton trapping rate for each packing motif, $k_{STE}$ is be calculated from $k_A = k_{TT} + k_{STE}$. Of particular interest is the solution to the coupled ODE system from Table S2 row(b).

$$n_{TT}(t) = n_o \left( \frac{k_{TT}}{k_{TT} - (k_{HT} + k_{T1})} \right) \left( e^{-(k_{HT}+k_{T1})t} - e^{-k_{TT}t} \right)$$

The above equation gives the triplet state population, $n_{TT}(T)$, as a rising and falling biexponential rate. The above equation fits the strongly rising polarization dynamics seen for resonant probe polarizations along the CT-axis projection (such as in Fig. 3). In the select cases where optical photons strongly assist TT formation, we instead use the generalized triexponential model shown in row(d) of Table S2.

### S.7 Using polarization to deconvolve the triplet pair dynamics

In general, all kinetic traces shown in this manuscript contain some residual singlet depletion dynamics regardless of probe energy or polarization. Fortunately, at the optimal TA ESA spectral peak in Fig. 5b, the probe polarization alone was sufficient to separate TT risings dynamics from the much weaker competing fall $S_1$ singlet contributions. However, Figure 7b demonstrates at the lower probe resonance of $E_{pr} = 1.15$ eV, that the data sets must be subtracted to recover the rising TT pair formation kinetics. This can be seen in 7b, where fitting the raw data and 'subtracted data' (open circles), yield similar rising exponential rates of 74±3 ps and 77±2 ps (biexponential fit, subtracted data). A similar analysis is done in Fig. 4a.

The anisotropy factor, $A_{pump}$ of the singlet depletion dynamics (and the $S_oS_1$ transition) is observed to be much smaller than the TT→$S_2$ (and the $S_oS_2$) anisotropy factors. To approximately correct for convolved singlet and TT dynamics at given probe polarization, we apply a $S_1$ decay subtraction method. Specifically, each kinetic trace is represented by ($\theta_{pump}/\theta_{probe}$)$_{conv}$, signifying the convolved $S_1$ and TT kinetics at the polarizations indicated. Considering the example of the diF-ADT TES kinetics shown in Fig. 7b, the resulting kinetic trace shown (red circles) is obtained by: $(90°/90°)_{TT} = (90°/90°)_{conv} - A_{probe} (90°/0°)_{conv}$. Here, the anisotropy factor is obtained by $A_{probe} = [(0/90°)_{t=0}/(0/0°)_{t=0}]$, i.e. the ratio of the max pump-probe amplitudes when only the probe polarization changes.

## S.8 Laser pump flux dependent TT-rise dynamics

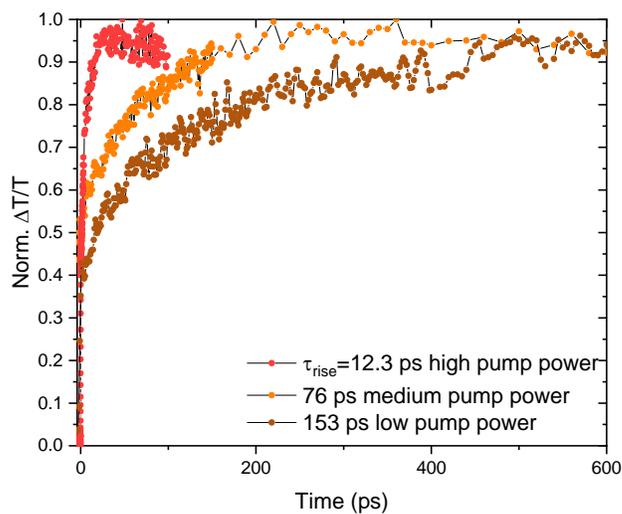

**Figure S8:** Comparison of diF-ADT TES when $S_1$ resonantly excited, and probed at 1.26 eV in 90° pump, 90° probe configuration. The exponential rise accelerates from 153 (brown) to 12.3 ps (red) at pump fluence increases.

## S.9 References